\documentclass{aa} 
\usepackage{graphics}

\newcommand{\nodata}{---}

\newcommand{\ngc}{NGC\,2359}
\newcommand{\hd}{HD\,56925}
\newcommand{\low}{$1 \rightarrow 0$}
\newcommand{\hig}{$2 \rightarrow 1$}
\newcommand{\kms}{km s$^{-1}$}

\newcommand{\hi}{{\sc Hi}}
\newcommand{\hii}{{\sc Hii}}
\newcommand{\htwo}{H$_2$}
\newcommand{\msun}{M$_\odot$}
\newcommand{\mlr}{M$_\odot$ yr$^{-1}$}
\newcommand{\tk}{T$_{\rm K}$}
\newcommand{\exc}{T$_{\rm ex}$}
\newcommand{\nh}{N$_{\rm HI}$}
\newcommand{\aj}{AJ}

\newcommand{\aap}{A\&A}

\newcommand{\aar}{A\&AR}
\newcommand{\acta}{Acta Astron.}
\newcommand{\apj}{ApJ}
\newcommand{\apjs}{ApJS}

\newcommand{\mnras}{MNRAS}

\begin{document}

\thesaurus{08            
  (08.09.2 HD 56925;     
   08.23.2;              
   08.23.3;              
   09.02.1;              
   09.09.1; NGC 2359     
   09.11.1)}             

\title{On the history of the interplay 
between \hd\ and \ngc}

\author{J.~R.~Rizzo\inst{1}\thanks{On leave of absence from 
Instituto Argentino de Radioastronom\'{\i}a, Argentina}
\and
J.~Mart\'{\i}n-Pintado\inst{1}
\and
J.~G.~Mangum\inst{2}}
\institute{Observatorio Astron\'omico Nacional, Aptdo. Correos 1143, 28800
Alcal\'a de Henares, Spain
\and 
NRAO, 949 N.\ Cherry Avenue, Tucson, AZ 85721}

\offprints{J.\ R.\ Rizzo}

\date{Received ; accepted } 

\titlerunning{CO and {\sc Hi} in NGC\ 2359}

\maketitle


\begin{abstract}
\ngc\ is an optical nebula excited by the powerful wind and the radiation
of the Wolf--Rayet star \hd. We have investigated the interaction 
between this massive star and the surrounding neutral gas by analyzing the
large-scale 21cm-\hi\ emission and by mapping the nebula in 
the J = \low\ and the J = \hig\ lines of CO. We found a
conspicuous ($70 \times 37$ pc) \hi\ shell, expanding at 12 \kms, likely
produced during the main-sequence phase of the star.
The molecular gas towards \ngc\ shows three velocity components. Two of 
these components, A1 and A2, have narrow linewidths (1--2 \kms) and radial
velocities of 35--38 and 64--68 \kms, respectively. The third component
is detected at radial velocities between 50 and 58 \kms\ and has a broader 
profile (up to 5.5 \kms). Furthermore, this component is morphologicaly 
related with the nebula and has a velocity gradient of a few \kms. We have 
also estimated the physical parameters of the molecular gas by means of
a LVG modelling of the CO emission. The gas projected onto the southern \hii\
region of the nebula has low CO column density and is rather hot, probably up
to 80\,K. Several profiles of the $^{13}$CO J = \low\ line near the peak of the
emission, together with a weak emission bridge between the broad and 
one of the narrow components (component A2), suggest the 
presence of a shock front acting in the southern part of the nebula. This
shock was likely produced in a previous RSG/LBV phase of \hd.

\keywords{stars: individual (\hd) -- stars: Wolf--Rayet -- stars: winds --
ISM: bubbles -- ISM: individual (\ngc) -- ISM: kinematics and dynamics}

\end{abstract}


\section{Introduction}

The different evolutive stages of massive stars have strong influence 
on their surroundings and, consequently, on the galactic evolution. 
During their lives, the massive stars inject large amounts of matter, 
energy and momentum into the interstellar medium (ISM). In the pre-main 
sequence, massive stars undergo a phase of very energetic mass loss giving 
rise to molecular outflows. UV radiation, expansion of \hii\ regions and 
fast stellar winds are present in the main sequence stage. In the 
short--lived red supergiant (RSG) and luminous blue variable (LBV) 
stages, the winds become slower and denser and ejections of up to some 
solar masses may be produced. The Wolf--Rayet (WR) stage, prior to SN 
explosions, is characterized by a copious mass loss (typically 
$10^{-5}$ \mlr) driven by a fast (about 2000 \kms) and chemically enriched 
stellar wind. Massive stars are therefore continuously heating, ionizing, 
shocking and blowing-up the ISM surrounding them. Garc\'{\i}a-Segura 
\& Mac Low (1995) have
modeled the evolution of the gas which surrounds massive stars, from the 
main sequence to the WR stage. They only considered simple models for 
the mass ejection in the form of stellar winds in the different stages 
of evolution and predict the presence of large scale structures around the 
massive stars. Both the blown-up gas and the ejecta form multiple shells 
as a consequence of the different shockfronts produced inside the gas.

During the main sequence phase, the gas around massive stars is ionized,
heated and highly evacuated by the fast, long-lived stellar wind and the
Lyman continuum radiation. A hot cavity is surrounded by an
expanding shell. The ionization front may get trapped within it, the shell may
start to recombine and become observable in the line radiation emitted by
neutral atoms or molecules. These large bubbles have been successfully
detected in the 21-cm line of atomic hydrogen (\cite{arn92}; \cite{riz94};
\cite{cap98}; \cite{riz98}; \cite{arn99}), in the form of huge 
shells of 30--70 pc, expanding at 6--20 \kms, with masses of
hundreds (or even thousands) of solar masses. Some of the large IRAS shells
found by Marston (1996) might also be related to this type of structures.

Once the massive star finishes the hydrogen burning in its nucleus and becomes
a RSG or LBV, its stellar wind suddenly becomes more dense, and a significant
amount of the stellar mass is deposited into the ISM at velocities around 20
\kms. Very probably, the formation of the optical ring nebulae starts in this
brief stage. When the star comes to a WR stage, the wind is accelerated to
hypersonic velocities and rapidly reaches the RSG/LBV wind. Many of the 
optical nebulae possess arcs and have high inhomogeneities in the 
abundances of heavy elements (\cite{chu83}, 1999).

However, the effects of the WR phase in the neutral (neither atomic nor
molecular) gas have not been easily detected. The structures formed during 
this phase would not be as large as those developed in the O-phase. Just in 
few cases, the neutral atomic hydrogen has been detected in interaction with 
ring nebulae (\cite{cap96}; \cite{arn96}). Vibrationally excited molecular
hydrogen has only been detected towards \ngc. The CO J = \low\ emission has
been observed around WR\,16 (\cite{mar99}) and in \ngc\ (\cite{sch81}).

\ngc\ is a very interesting object, and many aspects of which has been
thoroughly studied in the recent years. This optical nebula is excited by \hd\
(WR\,7 in the catalogue of van der Hutch et al.\ (1988), a WN4 star located in
the outer Milky Way. Its distance varies from 3.5 to 6.9 kpc depending on the
authors (see \cite{gou94} for a discussion). The mass--loss rate ($<7\times
10^{-5}$ \mlr) and the terminal wind velocity (1545 \kms) have been measured by
Schneps et al.\ (1981) and Rochowicz \& Niedzielski (1995), respectively.
\ngc, a prototypical wind-blown bubble (\cite{chu83}), is nearly spherical with
several small filaments inside. The nebula is not highly enriched of heavy
elements as a whole, with the exception of the filaments (\cite{est90}). This
may indicate that the nebula was largely produced in the RSG/LBV and the WR
stages of the exciting star. From the optical emission lines (\cite{gou83}) and
radio recombination lines (\cite{loc89}; \cite{fic91}), we know that the
ionized gas has several components at velocities between 30 and 70 \kms\ (all
velocities in this paper are referred to LSR).  An \hii\ region partially
surrounds the wind-blown bubble and is  apparently limited by obscuring
material. Schneps et al.\ (1981) have taken several profiles in the J = \low\
line of CO and detected the presence of three  clouds at 37, 54 and 67 \kms.
The presence of broad CO profiles at 54 \kms\ and  the morphology of the
emission, just following the south-eastern edge of the nebula, suggests that
this component has been interacting with the nebula. St-Louis et al.\ (1998)
have analyzed the \htwo\ emission in this region. They have found the 1--0 S(1)
line towards the southern border of the \hii\ region, but they could not
establish the nature of the excitation (fluorescence or  shocks) of the \htwo.
Cappa et al.\ (1999) made a complete map of the ionized component at the 1465
MHz continuum, and traced the \hi\ emission presumably connected with the
nebula. These \hi\ features are probably associated with the WR stage of \hd;
so far, the large atomic main-sequence bubble predicted by the theory has not
been yet reported.

The goal of this paper is to look into the history of the interaction of the
wind and the radiation of \hd\ and its surroundings. We have analyzed a public
\hi\ survey and studied for the first time the large scale interaction of 
\hd\ in the O-phase with the ambient ISM. Based on the mapping of the 
J = \low\ and J = \hig\ lines of CO over the whole optical nebula, we study in 
detail the
morphology and kinematics of the CO around \ngc. We also present several
$^{13}$CO profiles towards selected positions in order to determine the global
physical parameters of the molecular gas, such as kinetic temperature,
density and  opacity of the lines. This study has allowed us to shed some light
on the dominant excitation mechanism of the neutral gas in the WR stage.


\section{Observational material}

\subsection{The \hi\ survey}

We have analyzed a region of $6\degr \times 6\degr$ around \ngc\ using the
21 cm--\hi\ survey of Hartmann \& Burton (1997). This survey was made using the
25\,m radio telescope at Dwingeloo. The angular and velocity resolution of this
dataset are 36\arcmin\ and 1.03 \kms, respectively. The survey has been done
with a sampling of 30\arcmin\ in galactic coordinates and have a sensitivity
in brigthness temperature of 70 mK.

\subsection{CO observations}

The observations of the CO J = \low\ (115.271 GHz) and J = \hig\ (230.538 GHz)
lines and the J = \low\ (110.201 GHz) line of $^{13}$CO towards \ngc\ and its
surroundings were made during 2000 February, using the NRAO\footnote{The
National Radio Astronomy Observatory is operated by Associated Universities,
Inc.\ under cooperative agreement with the National Science Foundation.} 12\,m
radio  telescope at Kitt Peak. The half power beam width of the telescope was 
54\arcsec\ and 27\arcsec\ at the rest frequency of the J = \low\ and the J = \hig\
lines, respectively.

The data were acquired simultaneously with two independent SIS receiver
channels with orthogonal polarizations. The receiver temperatures were
250--300\,K and 500--600\,K for the J = \low\ and the J = \hig\ lines, respectively.
We used filterbanks (with 256 channels) and the hybrid correlator MAC (in
the 2 IF mode, in parallel, 3072 channels)  as backends. In the J = \low\
line, the filterbanks provided 100 and 250 kHz  of resolution, corresponding
to velocity resolutions of 0.26 and 0.65 \kms, respectively. For the same
line, the  MAC correlator was setup with a total usable bandwidth of 300
MHz and with channels of 195.3 kHz of resolution (0.51 \kms). In the J = \hig\
line, the filterbank with channel widths of 250 kHz (0.33 \kms) and 500 kHz 
(0.65 \kms) were used. We also used for this line the MAC correlator tuned
with an usable bandwidth of 600 MHz and frequency resolution of 390.6 kHz 
(0.51 \kms).

Mapping of the J = \low\ and J = \hig\ lines was made using the ``on-the-fly''
technique. We have obtained three individual maps in the J = \low\ line and 
six in the \hig\ line. These individual maps were later reduced
with AIPS and combined weighting them by their inverse square of
{\it rms}. For our investigation, we finally used the MAC data for
the J = \low\ line and the 500 kHz-filterbank data for the J = \hig\ line.
Typical final {\it rms} per grid  point was 120 mK and 80 mK in the J =
\low\ and J = \hig\ lines, respectively. We have also observed the J = \low\ line
of $^{13}$CO and C$^{18}$O at seven individual positions, with 7--10 minutes
of integration time achieving a {\it rms} noise of 30 mK.

The pointing was checked at the begining of each observing session with
five--point cross maps. The final data are pointed within 4'' of uncertainty.
The data were calibrated using the chopper wheel method. Final intensity scale
is in units of ${\rm T}_R^*$, e.g., corrected by atmospheric attenuation and
instrumental losses (\cite{kut81}). The calibration  was also checked off-line
by comparing individual spectra of standard sources.


\section{Results}

\subsection{Neutral hydrogen}

We have analyzed the \hi\ data in the velocity range 20--80 \kms, e.g., the
velocity range of the optical and the CO emission. According to most of
the galactic rotation models (for example \cite{fic89}), this broad range in
velocity is expected for the range of distances associated to \hd\
(\cite{gou94}). Within this velocity range, we looked for the structures 
connected to the evolution of \hd.

Fig.\ 1 depicts \hi\ column density (\nh) maps as a function of the radial 
velocity, indicated at the top left corner. The white circle near the 
center of the map sketchs the position and size of \ngc. Every map shown in 
Fig.\ 1 has been obtained by the integration of the brightness temperature 
over 2.1 \kms\ (2 channels) and assuming optically thin emission.

We have found an outstanding structure which remains coherent across tens of
velocity channels. This structure is a huge cavity surrounded by a set of
relative maxima in the \nh\ maps. The centroid of this cavity remains
approximately at the same point, and nearly off-centered from the nebula and
the WR star. The excentric position of the WR star with respect to the cavity
is also observed in almost all the WR bubbles (\cite{arn92}) and might be due
to the star motions and/or density variations in the original ISM
(\cite{mof98}). The structure reaches its maximum size at velocities between 61
and 65 \kms. For receding and approaching velocities with respect to this
velocity interval, the cavity diminishes its angular size, ending at $\sim$ 51
and 76 \kms\ with small relative maxima centered at the cavity.

The morphology of this structure has the typical ring--disk appearence of an
expanding bubble. Following the standard analysis of this type of shells, the
largest size of the cavity indicates the geometrical dimensions and its radial
velocity corresponds to approximately the systemic velocity of the gas before
the expansion. The full velocity range where the feature is observed gives a
lower limit of twice the expansion velocity. Table 1 summarizes the parameters
derived, as described above, for the expanding cavity. Following
Arnal et al.\ (1999), we have also derived lower and upper limits to the mass
of the \hi\ shell. These two values for the \hi\ mass are the missing mass into
the higher close contour and the neccesary mass to fill the cavity making it
indistinguishable of the nearly environs. All distance-dependent parameters
were calculated assuming a distance of 5 kpc.

This \hi\ shell is in geometrical coincidence with the IRAS shell reported
by Marston (1996). Furthermore, the mass of gas predicted from the IRAS data
falls within the \hi\ estimates presented in Table 1.

\subsection{Carbon monoxide}

\subsubsection{Morphology and kinematics}

The CO emission in the J = \low\ and J = \hig\ lines are observed in three
velocity components. The components with the lowest radial velocity (35--38
\kms) and with the highest radial velocity (64--69 \kms) show narrow profiles
of 1.5--2.5 \kms\ width. Hereafter we will refer to these two components as
Ambient 1 (A1) and Ambient 2 (A2), respectively. The third component, with
radial velocities of 50--58 \kms, shows broader profiles than the components A1
and A2, having linewidths of 4--5.5 \kms. We will refer to this component as
Broad component (B). Fig.\ 2 shows the global spatial distribution of the CO
J = \low\ (left panels) and J = \hig\ (right panels) integrated emission for
the three components. For comparison, the CO maps have been overlaid with a
J band image of \ngc, obtained from the {\it Digitized Sky Survey}. At the top
right corner of each panel, the velocity interval of integration is indicated.
The mapped regions in both spectral lines are also depicted. In the J = \low\
line, almost all the optical nebula and its surroundings has been covered by
our observations. In the J = \hig\ line, however, we limited the maps at the
east and the north-west because of our interest in studying in detail the
interaction of the nebula with the molecular gas.

The CO emission for components A1 and A2 appear mainly to the east and to the
south-east of the mapped regions, respectively. Although the spatial
distribution for both components are found close in projection to the optical
nebula, there is not a clear morphological correlation. Furthermore, there
are no significant changes in the line widths and in the peak velocities over
the regions where these components are observed. In contrast, the broad
component B is observed in clear correlation with the eastern and the southern
edges of \ngc, and with broader profiles, varying from 4 to 5.5 \kms\ wide.

The most intense CO emission of component B is located beyond the outer
edge of the \hii\ region in \ngc, mainly to the south-east part, but also
significant emission in the J = \hig\ line is detected towards locations
projected on the \hii\ region (middle right panel in Fig.\ 2). For a comparison
between the spatial distribution of the J = \low\ and \hig\ lines of CO
for the component B, the Fig.\ 3 shows and overlay of the emission
in the J = \low\ line (solid lines) and the J = \hig\ line (dashed lines)
convolved to the resolution of the J = \low\ line. This comparison shows that
both lines are well correlated at the most intense and the external part of
the nebula. However, we note again that the J = \hig\ line have significant
extended emission to the inner part of \ngc, especially into the southern
optical filament. This is clearly ilustrated in the left and right panels in
Fig.\ 3, where we show a sample of four CO J = \low\ and \hig\ line profiles
toward the positions sketched by squares in the central map. The lower left
spectra show the mean profiles around the most intense region, while  the
other three panel show the CO profiles in the periphery of the CO emission.
There is a clear trend in the profiles. While the J = \low\ line is stronger
than the J = \hig\ line toward the most intense zone of emission, the J =
\low\ line is much weaker than the J = \hig\ line or it is not detected in
positions closer to the \hii\ region. The \hig/\low\ intensity ratio varies
from 0.8--1 near the peak to 1.5--2 towards the positions near the \hii\
region. In contrast, the \hig/\low\ intensity ratio is nearly constant in
components A1 and A2, with values around 0.5--0.6.

While the components A1 and A2 do not have any remarkable kinematical
features, the component B have slightly different peak velocities as a
function of the position. Fig.\ 4 shows four position-velocity diagrams,
taken in the directions sketched in the map of the J = \hig\
emission at the top left corner of this figure. Each position-velocity
diagram is identified by the slice number. Slice 1 shows the
most intense part of component B at 53--54 \kms\ and the western border of the
component A2. Slice 1 also shows hints of a small velocity gradient in the
component B. This is more clearly depicted in the slice 2, where we find along
the central part a velocity shift of $\sim$ 2 \kms. Both slices 1 and 2
remarks the spatial coincidence between the eastern border of component B and
the western border of component A2. In fact, slice 3 shows the presence of a
weak (4 $\sigma$) ``bridge'' in the CO emission which connects the components
B and A2. This suggests a physical connection between both components.
Slice 3, parallel to the slice 2 but 1\arcmin\ to the south-west, does not
show signs of this ``bridge'', indicating the small size and weakness of this
bridge. The slice 4 show that the diffuse emission to the north have a radial
velocity slightly lower than that of the main part of the component B.

\subsubsection{$^{13}$CO J = \low\ emission in component B}

In order to study the excitation of the CO line in the component B, we
have observed seven positions in the $^{13}$CO and C$^{18}$O J = \low\ lines,
towards the southern part of \ngc. The location of the observed points are
shown in the Fig.\ 5a, superimpossed on a CO J = \hig\ map of the component B.
Five of these positions are located close to the slice 1. Figs.\ 5b, 5c and
5d plot three of these $^{13}$CO J = \low\ profiles (labeled with letters b, c
and d in the circles of Fig.\ 5a), together with the $^{12}$CO J = \low\
spectra of the same points. We find a significant variation in the
$^{12}$CO/$^{13}$CO ratio. This ratio is $\sim$ 6 for the CO emission peak
(Fig.\ 5b) and increases to values $>30$ elsewhere. 

Another interesting result is that the $^{12}$CO/$^{13}$CO ratio does not
only vary with position but also it changes with radial velocity within the
beam towards the CO emission peak (Fig.\ 5b). One can see that both lines peak
at different velocities. The $^{13}$CO line appears shifted by $\sim$ 1--2
\kms\ with respect to that of $^{12}$CO. This shift produces important
changes in the $^{12}$CO/$^{13}$CO ratio as a function of the radial velocity.
The $^{12}$CO/$^{13}$CO ratio is $\ga$ 30 at 50--53 \kms, and decreases to
$\sim$ 4--5 for radial velocities between 54 and 57 \kms. In view of the small
velocity gradient found in the component B, we do not think that this shift can
be due to a pointing error between the $^{12}$CO and $^{13}$CO data. Even in
the worst possible case, 10\arcsec\ of pointing difference, we find that this
ratio would change with radial velocity by at least a factor of 4 for
velocities below and above 53 \kms.


\section{Discussion}

\subsection{Origin of the \hi\ expanding shell}

The \hi\ structure found in this paper resembles the expanding shells often
associated with O stars (\cite{cap98}), WR stars (\cite{arn92}; \cite{arn99})
and OB associations (\cite{riz94}; \cite{riz98}). In the following we shall
explore if the \hi\ shell has been created by the stellar wind from \hd\ or
from its O--progenitor. Let us estimate, with a few assumptions, some important
parameters to distinguish between the two possibilities. We shall consider a
spherical expanding bubble, with a radius of 50 pc, an expansion velocity
of 12 \kms, and a mass of 1500 \msun. With these values, we estimate a
dynamical time for the shell of 2.3 Myr and a kinetic energy of $2.2\times
10^{48}$ erg. The dynamical time was computed as a mean value between the
adiabatic and the radiative case (\cite{mcc83}). This time is in fact an upper
limit, but a good approximation, to the age of the \hi\ structure
(\cite{dys89}). This age and the kinetic energy associated to the \hi\ shell
clearly indicate that its origin is due to the main-sequence phase of \hd.
Indeed, the most massive stars, greater than 25 \msun, remains a few million
years at the main sequence (\cite{mae94}) and creates interstellar bubbles of
30--60 pc in this stage (\cite{wea77}; \cite{gar95}). It is thought that WR's
are descendant of these massive stars. On the other hand, the RSG, LBV or WR
phases are short-lived, probably $10^4 - 10^5$ yr (\cite{mae94}; \cite{lan94}).
We therefore conclude that only the O progenitor of \hd\ had sufficient time to
develop the \hi\ expanding shell reported in this paper. A similar conclusion
was reached by Marston (1996) concerning to the large IRAS shell found around
several WR stars, including the one linked to \hd.

By adopting typical values for the mass-loss rate of $10^{-6}$ \mlr\ and a
wind velocity of 1000 \kms\ for the O--progenitor star, we estimate that this
star deposited into the ISM during 2.3 Myr a total energy of $\sim 2.3 \times
10^{49}$ erg in the form of stellar wind. If this star has blown up the shell
which we are observed in \hi, it implies a kinematical efficiency of nearly
10\%. Although the uncertainties in these computations are large and hard to
be estimated, this value for the efficiency is in good agreement with the
models that predict this type of structures (\cite{wea77}; \cite{van86}) and
reinforces our hypothesis of a main sequence origin of the \hi\ shell.

\subsection{Physical parameters of the molecular gas}

We have estimated the physical parameters of the molecular gas from the CO
line emission by modelling the excitation of the CO lines using the {\it
Large Velocity Gradient} (LVG) approximation. For comparison with the model, 
we have
smoothed the J = \hig\ line data to the J = \low\ resolution. For component
B, the $^{13}$CO data was also taken into account in order to estimate the
opacity of the $^{12}$CO lines. Six different components have been defined.
The first three are the ambient components A1 and A2 and the most
extended emission of the component B (hereafter B$_{\rm all}$). The fourth
component considered is the one projected onto the southern \hii\ region, named
as B$_{\rm HII}$. Finally, we also considered the region around the CO maxima
in component B, where the $^{13}$CO behaviour is striking. We have subdivided
this component in the two velocity ranges having the greatest difference in
the $^{12}$CO/$^{13}$CO ratio (see Fig.\ 5b), named as B$_{\rm max}^{blue}$
(for the velocity interval 50--54 \kms) and B$_{\rm max}^{red}$ (for the
velocity interval 54--58 \kms). The results derived from the LVG analysis are
shown in Table 2. This table gives estimates for the excitation temperature
(\exc), \htwo\ density (n(\htwo)), CO column density (N(CO)), the opacities
of the CO J = \low, CO J = \hig\ and $^{13}$CO J = \low\ lines, and the total
mass of the regions (m(\htwo)), derived from the two line
ratios (second and third columns) and the $^{13}$CO intensities.
A kinetic temperature (\tk) of 10\,K was assumed for all the
regions but B$_{\rm HII}$. The phyisical conditions in
B$_{\rm HII}$ seem to be very different from those of the other regions. The
large ratio between the J = \hig\ and the J = \low\ lines is explained by
optically thin emission with relatively high \tk. The most likely \tk\ is
about 80\,K and we present in Table 2 the results derived for this \tk.
Components A1 and A2 are subtermally excited (\exc\ $<$ \tk), with typical
\exc\ of 5--6\,K. We obtain higher excitation temperatures in component B.
For B$_{\rm max}^{blue}$ and B$_{\rm max}^{red}$, we derived \exc\ similar to
\tk, indicating that the lines are thermalized, and densities significantly
higher than for A1 and A2. 

For B$_{\rm max}^{red}$ we did not apply the LVG method since this region has
a very low value of the $^{12}$CO/$^{13}$CO ratio. Such low value for this
ratio usually indicates the presence of a region of very high opactiy in the
$^{12}$CO lines. Since we have found this low ratio toward only one observed
position, we conclude that the emiting region corresponding to 
B$_{\rm max}^{red}$ is unresolved by the J = \low\ beam. To derive the
physical properties in this component, we have assumed an optically thin
emission for the $^{13}$CO and optically thick emission of the $^{12}$CO. For
a $^{12}$CO/$^{13}$CO isotopic ratio of 60 (\cite{wil92}), we derive
an optical depth of 12 for the J = \low\ line of $^{12}$CO. From the
$^{12}$CO J = \low\ line intensity and assuming that the emission in this line
is thermalized to the \tk\ of 10\,K, we derive a beam filling factor for
B$_{\rm max}^{red}$ between 0.1 and 0.2. With the optical depth we have
estimated N(CO). Adopting a beam filling factor of 0.15 and a fractional
abundance for CO of $10^{-4}$ we have also estimated n(\htwo). Although our
estimates of the \htwo\ density can be uncertain because of the unknown \tk,
the large differences in the \htwo\ density between B$_{\rm max}^{red}$ and
the other components indicate that a rather small region has a density larger
than other components.

\subsection{Molecular gas interacting with the nebula and the WR star}

In Sect.~3.2 we have discussed the morphological and kinematical arguments
for the relationship between the different CO components and \ngc. In the
previous section, we have derived their physical conditions. In this
section, we shall consider our observational findings together with those
already published, in order to address the nature and the origin of the
interaction of the molecular gas with the nebula and its exciting WR star. The
component A1 is located mainly to the NE of the map, while the component A2 is
present to the SE of the nebula. Both components are narrow (2 \kms) and do
not show any morphological or kinematical structures that indicate any
disturbance produced by the nebula or the WR star. In contrast, the component B
has some kind of ubiquity in this field, in agreement with optical emission
lines like the [\ion{O}{3}]. It appears as a broad component (4.5--5 \kms)
following the east and the south border of the \hii\ region which surrounds
the nearly-spherical bubble. We noted that the CO emission is stronger {\it
outside} the optical border of the \hii\ region, but significant emission is
also present in projection {\it onto} the \hii\ region. We think that
the component B is the only component which undoubtely interacts with \ngc,
because of its morphology, the large width of the lines, the higher
temperature and the striking kinematics.

As already mentioned, the eastern part and the southern border of the
\hii\ region are clearly bounded by the most intense component B. Furthermore,
it is also striking the presence of faint CO emission perfectly correlated
with the \hii\ region. These results are also in agreement with the
vibrationally excited \htwo\ emission observed at 2.2$\mu$m (\cite{nic98}).
The CO emission is dense and more opaque outside the \hii\ region, while onto
the \hii\ region it is less dense and has a higher temperature (up to 80\,K).
This temperature is compatible with the dust temperature inferred by
Marston (1991) in the same region.

At this point we wish to consider the $^{13}$CO behaviour in the southern part
of the component B. We have only detected this molecule near the peak of the
CO emission. This indicates that this is a {\it unique} region where the
$^{12}$CO/$^{13}$CO changes appreciably, most likely due to an increase of the
opacity in the $^{12}$CO lines. Only 1\arcmin\ to the east of this position
(essentially one beam) there is no significant emission in $^{12}$CO from the
component B, just at the western border of the component A2. This spatial
anti-correlation between components A2 and B is also present toward the
southern interface. Furthermore, there is a bridge in the CO emission which
seems to connect both components within a small area (see slice 3 of Fig.\ 4),
suggesting that both components are related. The CO in the \hii\ region might
be heated by the WR radiation field (\cite{nic98}), but the interface between
the components A2 and B may not be excited by the same mechanism.

All these observational facts can be explained in a scenario in which a shock
of 10--14 \kms, driven by the expanding hot bubble, impacts on the component
A2. As a consequence of the shock, the gas was accelerated up to the
velocities of the component B. The large width of the component B when
compared with the component A2 can be understood by means of an increase
of the turbulent motion due to the energy injected by the shock. Furthermore,
the systemic velocity of the \hi\ shell found at larger scales (presumably near
the rest velocity of the gas) is also similar to the velocity of the component
A2, supporting this scenario. The shock heats the gas and forms a {\it
thin layer} of compressed material with large \htwo\ density and with high
opacity in the $^{12}$CO. This thin shocked layer would remain unresolved by
the beam. The kinematics shown in Fig.\ 4 and the low $^{12}$CO/$^{13}$CO
ratio at 53--56 \kms\ towards the CO peak are very suggestive of this thin
shocked layer. Indeed, the large optical depth of the CO lines can only
be explained if this emission arises from a very small region unresolved by
the beam, probably less than 0.2 pc wide. Furthermore, the large CO 
column density of B$_{\rm max}^{red}$ as compared with B$_{\rm max}^{blue}$ would 
indicate large densities for these radial velocities (up to several 
10$^4$ cm$^{-3}$). Obviously, in the scenario of the shocked gas responsible
for the component B, B$_{\rm max}^{red}$ would represent the material which
has been recently shocked. High angular resolution observations are needed 
to disclose this particular region and to check our interpretation.

Although our hypothesis seems reasonable, it should be pointed out that this
shocked region is small and it can not be spatially resolved in our data.
Another crucial point is the weakness of the ``bridge'' which connects the
components A2 and B. Higher angular resolution and more sensitive
observations are needed to confirm the presence of the shocked layer and to
fully study its properties. If this is the case, \ngc\ would have the first
direct evidence of shocked molecular gas in a WR environment. The O-stage of
\hd\ have created in the ISM the \hi\ shell reported in this paper and the
IRAS shell reported by Marston (1996). In consequence, the optical nebula and
its surroundings, where we find most of the CO emission, have a more recent
origin than the large scale \hi\ shell. The \hii\ region has a size that
indicate a dynamical age of less than 10$^5$ yr. This age is comparable with
the WR or a previous phase, such as LBV or RSG (\cite{mae94}). On the other
hand, many optical shells of enriched material are found close to the WR star
(\cite{est90}). This leads us to think that the material now interacting with
the \hii\ region was ejected in previous episodes of RSG or LBV phase.
Chemical studies using other molecular line transitions, especially
those capable of tracing the molecular gas enriched by these phases, will
definitively confirm or discard these ideas.

\section{Conclusions}

We have analyzed the large-scale 21 cm-\hi\ emission in a $6\degr \times
6\degr$ field around the nebula \ngc, and found a large expanding bubble
roughly centered at the nebula and its exciting star, the Wolf-Rayet \hd.
The systemic velocity of the bubble is $\sim$ 64 \kms\ and it is expanding at
12 \kms. At an assumed distance of 5 kpc, the bubble has a size of $70 \times
37$ pc and a \hi\ mass between 700 and 2400 \msun. By simple estimates of
energetics, ages and stellar wind parameters, we think that this feature is a
wind-blown bubble produced by the O-progenitor of the WR star. The presence of
a nearly spatially-coincident IRAS shell reinforces our hypothesis.

We have also mapped the CO J = \low\ and \hig\ emission adjacent to \ngc\ and
found three CO components in this region. Two of these components, with radial
velocities of 34 (component A1) and 67 (component A2) \kms, have
narrow profiles (up to 2 \kms) and do not show any morphological or kinematical
effects which indicate the disturbance by the nebula or the WR star. However,
the third component (the component B, emitting at 54 \kms), clearly bounds
\ngc\ at its southern and eastern border. The profiles are broad, with
linewidths of 4--5.5 \kms. A velocity gradient of a few \kms\ is noted
towards the south and south-eastern interface with the nebula. A weak
``bridge'' in velocity of small angular extension seems to connect this
component with the component B2 at the southern part.

We have observed the $^{13}$CO J = \low\ line at a few points in the region
where the component B is detected. We have found a striking variation of the
$^{12}$CO/$^{13}$CO ratio both in position and in velocity. This ratio seems
to decrease near the peak of the CO emission, for velocities higher 
than 53 \kms. This particular emiting region is not resolved by the beam, 
and is probably 0.2 pc wide.

From the observational data, we have estimated the physical conditions for
each CO component using the LVG approximation. The highest excitation
temperature and density are found in the component B. It is remarkable that
high temperatures (up to 80\,K) are derived for the CO emission projected
onto the \hii\ region .

In view of the kinematics, the morphology and the physical properties of the
CO, we think that the component A2 was shocked and accelerated by the
expanding bubble to reach the radial velocities of the component B. The shock
front is still acting at the southern part of the region. The origin of such
shock might be related to the WR phase of \hd\ or, more probably, to previous
episodes of RSG of LBV. From the $^{12}$CO/$^{13}$CO ratio we infer the
presence of a thin layer of dense gas which represent the more recently
shocked material. Our data, however, cannot resolve spatially the region
where a thin layer of shocked material is expected. CO observations with
better resolution and sensitivity is needed to disclose the main features of
the region.



\clearpage

\begin{table}
\caption[]{Main parameters of the \hi\ bubble}
\begin{tabular}{l@{\quad}l@{\qquad}r}
\hline
\noalign{\smallskip}
Parameter & Unit & Value \\
\noalign{\smallskip}
\hline
\noalign{\smallskip}
Systemic velocity	& \kms\		& 64\\
Velocity range		& \kms\ 	& 51 to 76\\
Major axis$^{\mathrm{a}}$	& pc	& 70\\
Minor axis$^{\mathrm{a}}$	& pc	& 37\\
\hi\ mass, lower limit$^{\mathrm{a}}$	& \msun\	& 700\\
\hi\ mass, upper limit$^{\mathrm{a}}$	& \msun\	& 2400\\
\noalign{\smallskip}
\hline
\noalign{\smallskip}
$^{\mathrm{a}}$ For an assumed distance of 5 kpc.
\end{tabular}

\end{table}



\begin{table*}
\caption[]{Physical parameters of the CO emiting regions}
\begin{tabular}{l@{\quad}ccccccccc}
\hline
\noalign{\smallskip}
Component & 
$\frac{2\rightarrow 1}{1\rightarrow 0}$ & 
$\frac{^{12}{\rm CO}}{^{13}{\rm CO}}$ & 
T$_{ex}$ & n(\htwo) & 
N(CO) & $\tau_{10}$ & $\tau_{21}$ 
& $\tau(^{13}{\rm CO}$) & m(\htwo)\\

& & & K & $10^3$ cm$^{-3}$ & $10^{16}$ cm$^{-2}$ & & & & \msun\\

\noalign{\smallskip}
\hline
\noalign{\smallskip}

A1	    & 0.5 & \nodata &  5 &  0.6 & 1.0 & 0.6  & 0.6  & \nodata  & 96\\

A2          & 0.5 & \nodata &  6 &  1.0 & 0.4 & 0.8  & 0.8  & \nodata  & 28\\

B$_{\rm all}$ & 0.8 & \nodata & 8 & 1.8 & 0.5 & 0.3  & 0.4  & \nodata  & 160\\

B$_{\rm max}^{blue}$ & 1.0 & 30 & 10 & 5.0 & 1.6 & 0.6 & 0.8 & $<0.1$ & 10\\

B$_{\rm max}^{red}$  & 1.0 & 5 & 10 & 58 & 21. & 12 &\nodata& $\sim$ 0.1 & 120\\

B$_{\rm HII}$ & 2.0 & $\ga$ 30 & $\ga$ 80 & $<1.6$ & 0.1 &$<$0.1& 0.2 & $\leq$
0.1 & $\la$ 8\\

\noalign{\smallskip}
\hline
\noalign{\smallskip}
\end{tabular}
\\

The LVG approximation was used in all cases but B$_{\rm max}^{red}$. For this 
particular region, we have \\
estimated the size of the emiting region and taken into account the beam 
filling factor. \\
All masses were computed by assuming a distance of 5 kpc.

\end{table*}

\clearpage
\begin{figure*}
{\large Figure captions}\\
\caption{Distribution of the 21 cm--\hi\ column density in a large
region around \ngc, obtained from the survey of Dwingeloo (\cite{har97}).
Every map has been constructed integrating over 2.1 \kms (2 channels), being
the central velocity of integration indicated at the top right corner.
Contour levels are 3 to 18 times 10$^{19}$ cm$^{-2}$. The small circle near
the centre indicates the position and size of \ngc. The Dwingeloo beam is also
plotted in the bottom left map}
\end{figure*}

\begin{figure*}
\caption{CO emission in the field of \ngc. The three maps at
left correspond to the J = \low\ line, while the three maps at right
correspond to the J = \hig\ line. The three principal components detected are
named as Ambient 1 (A1), Broad (B) and Ambient 2 (A2), indicated at the top
right corner of every map, together with the velocity of integration in \kms.
Beams are also depicted at the bottom maps. The polygons drawn enclose the
mapped regions in both spectral lines. Contour levels are (1, 2, 4, 6, 9, 12,
16, 20, 25 and 30) times 0.2 K \kms\ (components A1 and A2, J = \low\ line),
0.3 K \kms\ (A1 and A2, J = \hig\ line) or 0.5 K \kms\ (component B, both
lines). The optical image of \ngc\ has been  obtained from the {\it Digitized
Sky Survey}}
\end{figure*}

\begin{figure*}
\caption{Comparison of the emission between the two observed
transitions of CO in the B component. In the central panel, solid lines
represent the J = \low\ intensity, while the dashed ones represent the J
= \hig\ intensity, convolved to the J = \low\ resolution. Contour levels are 
(1, 2, 4, 6, 9, 12 and 16) times 0.5 K \kms. Four selected positions are
indicated by squares, and their spectra  plotted at left and right panels.
Every panel displays the J = \hig\ spectrum  above the J = \low\ spectrum}
\end{figure*}

\begin{figure*}
\caption{Four strips in position for the J = \hig\ emission of
CO. The location of the slices are sketched by right lines in the small map at
the top left of the figure. The numbers indicate the slice number and the
origin of the angular distance scale in the four slices. Contour levels are 1
to 8 times 0.2\,K and 5 to 12 times 0.4\,K. Resolution in both position and
velocity are plotted in slice 3. Arrows indicate the location of the
components A2 and B}
\end{figure*}

\begin{figure*}
\caption{Comparison of the $^{12}$CO and $^{13}$CO \low\ lines in
three selected positions, indicated at the upper left of every figure. Figure
{\bf a} reproduces part of the J = \hig\ map of component B, superimpossed to
the positions where the $^{13}$CO J = \low\ line was observed. Letters b, c
and d inside three of these circles indicate the position of the profiles
shown in panels {\bf b}, {\bf c} and {\bf d}, respectively. Different scale
was used for both isotopes. It is remarkable the variations of the
$^{12}$CO/$^{13}$CO ratio from one point to  another, and the different
velocity peaks on both isotopes of panel {\bf b}}
\end{figure*}

\end{document}